\documentstyle[11pt,newpasp,twoside]{article}
\newcommand{\PSbox}[3]{\mbox{\rule{0in}{#3}\includegraphics{#1}\hspace{#2}}}

\begin{document}

\title{The Nature of Lyman Break Galaxies in Cosmological Hydrodynamic Simulations}
\author{Romeel Dav\'e}
\affil{Princeton University Observatory, Princeton, NJ 08544}
\author{Jeffrey P. Gardner}
\affil{Dept. of Astronomy, University of Washington, Seattle, WA 98195}
\author{Lars Hernquist}
\affil{Harvard-Smithsonian Center for Astrophysics, Cambridge, MA, 02138}
\author{Neal Katz}
\affil{Dept. of Astronomy, University of Massachusetts, Amherst, MA, 01003}
\author{David H. Weinberg}
\affil{Dept. of Astronomy, Ohio State University, Columbus, OH 43210}

\begin{abstract}
What type of objects are being detected as $z\sim 3$ ``Lyman break
galaxies"?  Are they predominantly the most massive galaxies at that
epoch, or are many of them smaller galaxies undergoing a short-lived
burst of merger-induced star formation?  We attempt to address this
question using high-resolution cosmological hydrodynamic simulations
including star formation and feedback.  Our $\Lambda$CDM simulation,
together with Bruzual-Charlot population synthesis models, reproduces
the observed number density and luminosity function of Lyman break
galaxies when dust is incorporated.  The inclusion of dust is crucial
for this agreement.  In our simulation, these galaxies are
predominantly the most massive objects at this epoch, and have a
significant population of older stars.  Nevertheless, it is possible
that our simulations lack the resolution and requisite physics to
produce starbursts, despite having a physical resolution of $\la
700$~pc at $z=3$.  Thus we cannot rule out merger-induced starburst
galaxies also contributing to the observed population of high-redshift
objects.
\end{abstract}


\section{Introduction}

The detection of large numbers of high-redshift galaxies using the
Lyman break technique has greatly furthered our understanding of early
galaxy formation.  A variety of arguments, from clustering[1] to
semi-analytic modeling[2] to N-body simulations[3], suggest that these
Lyman break galaxies (LBGs) form in highly biased, rare density peaks
in the early universe.  However, the nature of these galaxies remains
controversial.  Are they the most massive galaxy contained in these
peaks, having quiescently formed stars for some time[4]?  Or are they
smaller galaxies residing in large potential wells that are undergoing
an short-lived merger-induced starburst[5]?  The key to answering this
question is to determine the mass of the underlying galaxy.  This may
be done observationally[6] or by modeling processes of galaxy
formation[7].  So far, only N-body and semi-analytic techniques have
been applied, and the results vary, depending primarily on what is
assumed for merger-induced starbursts.  In principle, hydrodynamic
simulations of galaxy formation including star formation, together with
population synthesis models, can directly address these questions
within a given cosmology.  That is what we investigate in these
proceedings.

\section{Simulation and Analysis}

We simulate a random 11.111$h^{-1}$Mpc cube in a $\Lambda$CDM universe,
with $\Omega_m=0.4$, $\Omega_\Lambda=0.6$, $H_0=65$, $n=0.95$, and
$\Omega_b=0.02h^{-2}$.  We use Parallel TreeSPH to advance $128^3$ gas
and $128^3$ dark matter particles from $z=49$ to $z=3$.  Our spatial
resolution is $1.7h^{-1}$ comoving kpc (equivalent Plummer softening),
implying that at $z=3$ our physical resolution is $\sim 640$pc.
Our mass resolution is $m_{SPH}=1.3\times 10^7 M_\odot$ and
$m_{dark}=1\times 10^8 M_\odot$.  Using a 60-particle criterion for our
simulated galaxy completeness limit[8] implies that we are resolving
most galaxies with $M_{baryonic}\ga 8\times 10^8 M_\odot$.

We include star formation and thermal feedback[9].  At $z=3$, we
identify galaxies using Spline Kernel Interpolative DENMAX (SKID), and
compile a list of star formation events in each galaxy.  Since gas is
gradually converted into stars in each SPH particle, a given particle
can have up to 20 star formation events.  We treat each event as an
instantaneous single-burst population using Bruzual \& Charlot's
GISSEL98[10], assuming a Scalo IMF with $Z=0.4Z_\odot$\footnote{Using a
Salpeter or Miller-Scalo IMF results in more LBGs.  However, dust plays
a larger role in determining galaxy properties.}.  We sum the spectra
for all events in a galaxy to produce its rest-frame spectrum at
$z=3$.  We apply a correction for dust absorption using a galactic
extinction law[11] with $A_V=1.0$.  We then redshift the spectra to
$z=0$ and apply $U_nGR$ filter functions[12] to obtain the observed
broad-band colors for our simulated galaxy population.  Note that no
K-correction is necessary since we redshift the spectrum prior to
applying the filters.

\section{The Simulated Lyman Break Galaxy Population}

Our simulation produces 1238 galaxies at $z=3$.  Figure~1 shows the
luminosity functions $\Phi$ in $R$ (solid histogram), $G$ (dotted line) and
$U_n$ (dashed line) of these galaxies.  Note that $\Phi(U_n)$ is shown
without any attenuation due to HI along the line of sight.  The left
and right panels show $\Phi$ without and with dust, respectively.  The
turnover above $R\ga 28$ is likely due to resolution effects, while the
lack of galaxies with $R\la 24$ is due to our small volume.  Between
these values, our luminosity function (with dust) is in rough agreement
with the observed $R$-band luminosity function~[13], shown as the solid
curve down to $R=27$ (the current observational limit), although
somewhat steeper.  In reality, there is probably a range of dust
extinctions, and this will tend to flatten $\Phi$.

\PSbox{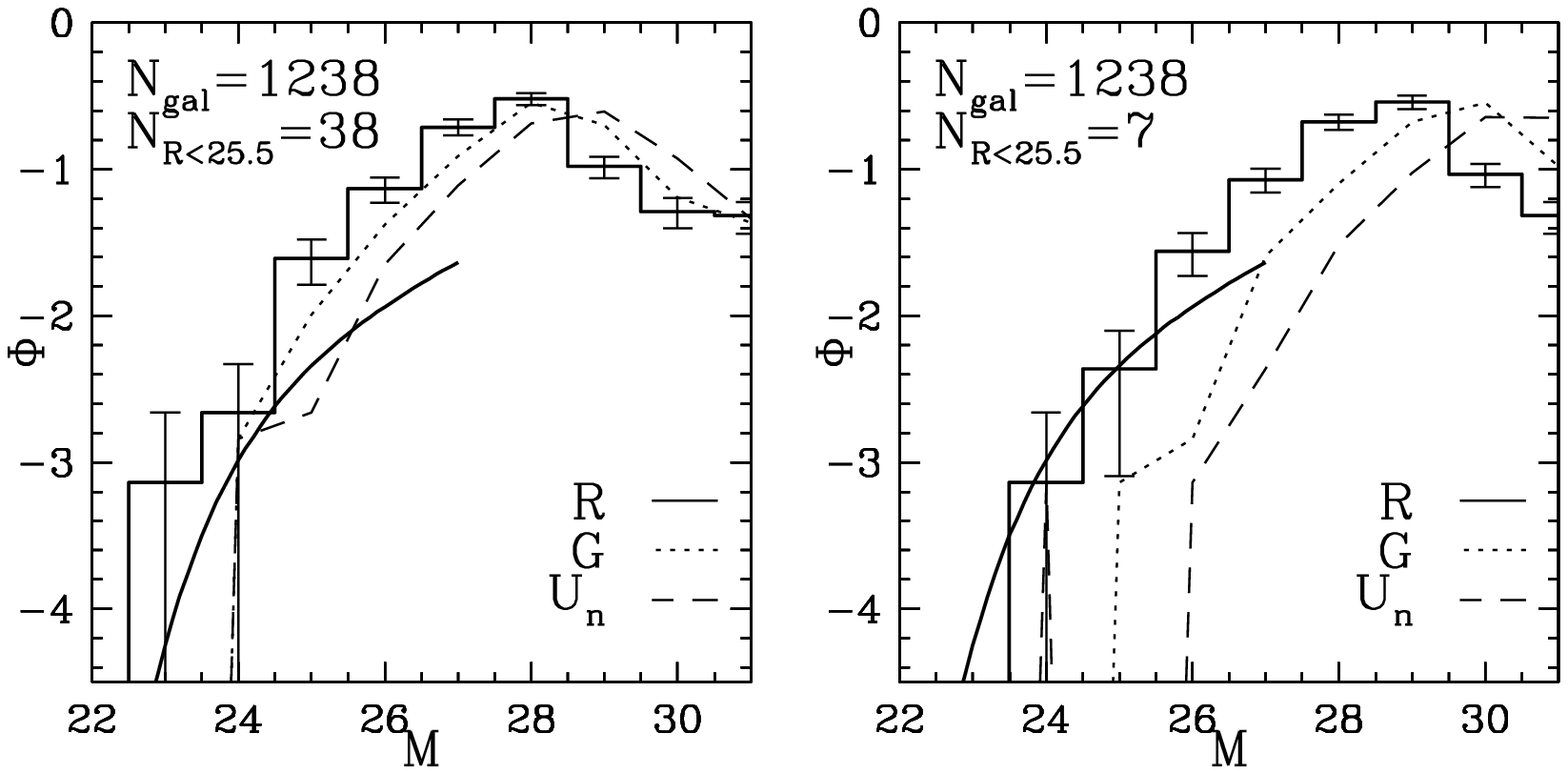 angle=0 voffset=-250 hoffset=0 vscale=51 hscale=51}{3.0in}{1.8in}
{\\\small Figure 1:
Luminosity function of high-redshift galaxies in $U_n$, $G$ and $R$,
without (left panel) and with (right panel) dust.
\label{fig: lumfcn} }
\vskip0.1in

The number of Lyman break galaxies expected for this cosmology and
volume is $\sim 7$~[1,13],  though this number could be higher due to
source confusion~[15].  With dust included, we produce 7 galaxies with
$R<25.5$, of which 6 satisfy the LBG color selection, in reasonable
agreement with observations.  Without dust, there are 38.  Not
surprisingly, the number density of simulated LBGs is highly sensitive
to the amount of dust included, and undoubtedly to the type and
distribution of dust as well.

\PSbox{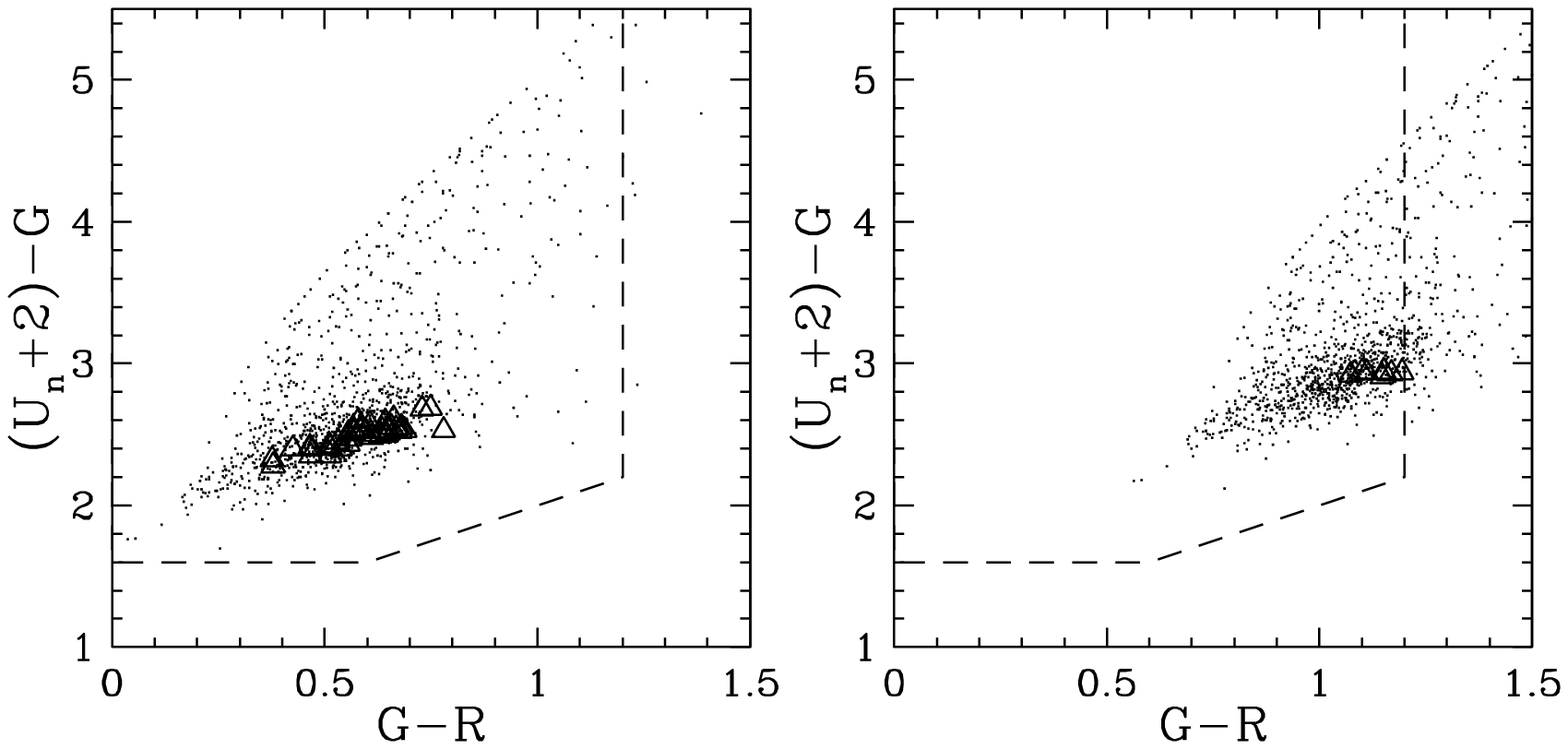 angle=0 voffset=-250 hoffset=0 vscale=51 hscale=51}{3.0in}{1.8in}
{\\\small Figure 2:
$(U_n+2)-G$ vs. $G-R$ of simulated galaxies, without (left panel) and with
(right panel) dust.  Triangles have $R<25.5$, dots have $R>25.5$.
\label{fig: colcol} }
\vskip0.1in

Color selection is at the heart of the Lyman break technique.  In
Figure~2 we show $U_n-G$ vs. $G-R$ plots of our simulated galaxies,
with an arbitrary two magnitudes of extinction added to $U_n$ to
crudely mimic intervening HI absorption.  Triangles represent
galaxies with $R<25.5$, and dots are the remaining galaxies.  The Lyman
break color selection is up and to the left of the dashed boundary.
Left and right panels show without and with dust, respectively.  Dust
moves galaxies to higher $G-R$, and somewhat higher $U_n-G$.  Most
galaxies at $z=3$ fall within the color selection, but significantly
more dust would move the bright galaxies outside the $G-R<1.2$
criterion.

\PSbox{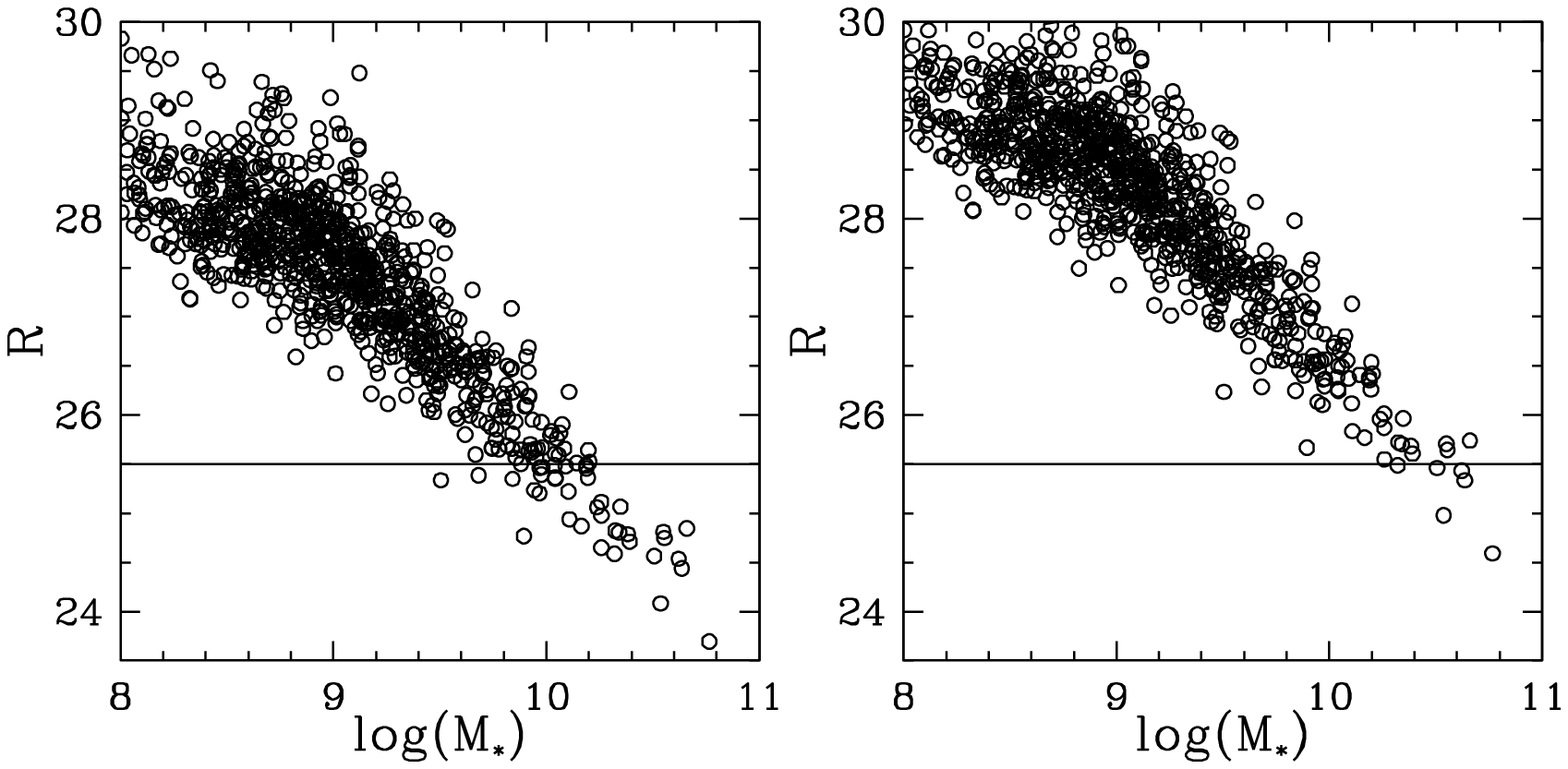 angle=0 voffset=-250 hoffset=0 vscale=51 hscale=51}{3.0in}{1.8in}
{\\\small Figure 3:
Stellar mass vs. $R$, without dust (left panel) and with dust (right panel).
\label{fig: MvsL} }
\vskip0.1in
We now investigate the mass of simulated LBGs.  Figure~3 shows the
stellar mass vs. $R$-band magnitude.  The horizontal line demarcates
$R=25.5$, the magnitude limit of the observed LBG sample.  While there
is some scatter, the clear trend is that the brightest objects are also
the most massive ones.  The scatter increases to smaller masses, and is
slightly larger in $G$ and $U_n$, but our simulations indicate that
LBGs are the most massive galaxies at $z=3$.

\PSbox{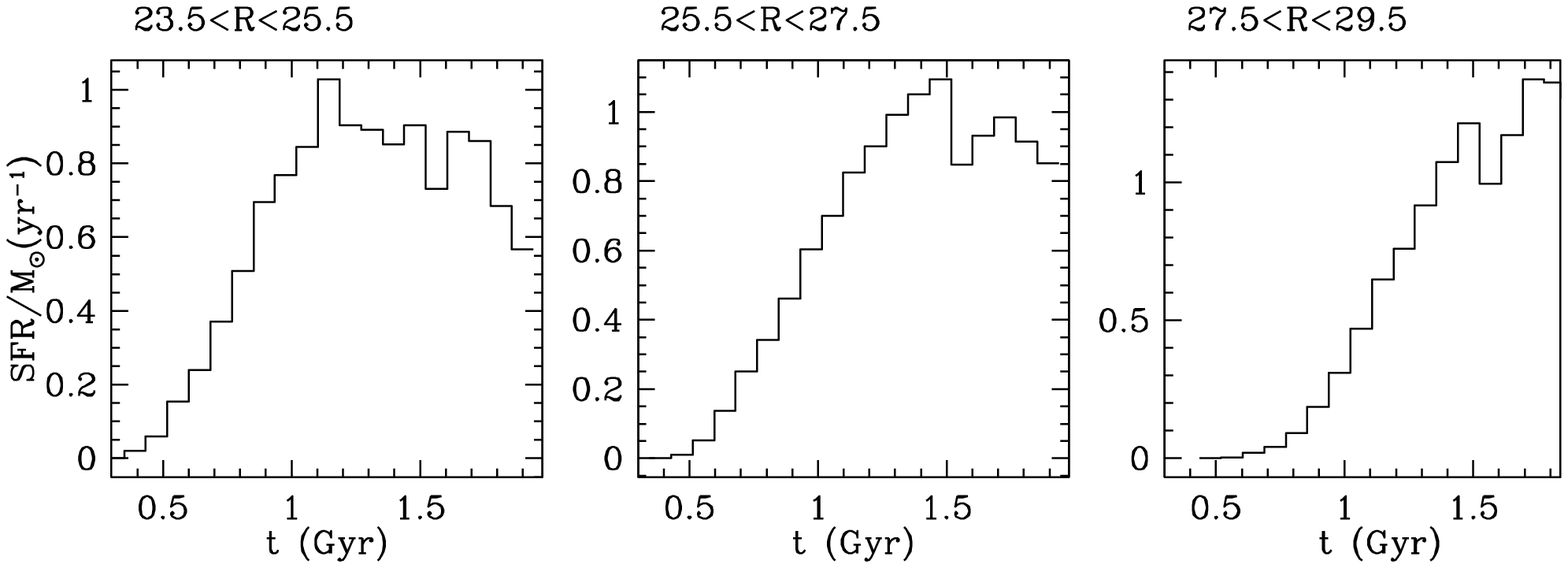 angle=0 voffset=-265 hoffset=0 vscale=51 hscale=51}{3.0in}{1.5in}
{\\\small Figure 4:
Star formation rate per unit stellar mass as a function of time (age of
the universe), in three different $R$-band magnitude ranges.
\label{fig: sfhist} }
\vskip0.1in
Figure~4 shows the evolution of the star formation rate per solar mass
of stars for three galaxy samples: $R<25.5$ (left panel), $25.5<R<27.5$
(middle panel), and $27.5<R<29.5$ (right panel).  The brightest
galaxies have been forming stars the longest, typically for over a Gyr
by $z=3$.  Thus they contain a significant older stellar population.
Fainter (and smaller) galaxies have formed the bulk of their stars 
more recently.

\section{Conclusions}

Our simulation roughly reproduces the number density and luminosity
function of LBGs for a reasonable value of dust extinction.  It
suggests that LBGs are the most massive objects at $z\sim 3$, and that
they contain a significant older stellar population.

While this simulation puts forth a consistent picture for the nature of
LBGs, we cannot rule out the aforementioned alternative scenario that
LBGs are smaller starbursting galaxies.  The reason is that starburst
regions are typically a few hundred parsecs across, and therefore below
our resolution.  The star formation rate in our simulations is tied
primarily to the local density (using a Schmidt Law) which is limited
by resolution.  Thus we do not effectively mimic ``starbursts" as would
occur in a much higher resolution merger simulation[14].  Conversely,
some semi-analytic models insert such starburst behavior explicitly, so
it is not surprising that they obtain different results.

At present, it is not feasible to run simulations of sufficient
resolution to resolve starbursts while still modeling a random
cosmological volume.  Furthermore, starbursts are likely to be governed
by many other physical processes that we are only crudely modeling at
present, such as feedback and ionization.  Thus we cannot rule out the
possibility that smaller starbursting systems also contribute to the
observed Lyman break galaxy population.  Nevertheless, our simulation
with reasonable physical parameters is able to reproduce the basic
observed properties of this population without including such objects.

\acknowledgments
Thanks to Kurt Adelberger for the $U_nGR$ filter response functions.
Thanks to Stephane Charlot for GISSEL98.  Thanks to Rachel Somerville
and Tsafrir Kolatt for helpful comments.  The simulation was run on the
NCSA Origin 2000.

\end{document}